\documentclass{article}
\usepackage[psamsfonts]{amssymb}
\usepackage{amsmath}
\usepackage{cmmib57}
\usepackage[cmtip,arrow]{xy}
\usepackage{pb-diagram,pb-xy}
\newcommand{\bPf}{\par\vspace*{-4pt}\indent{\sc Proof.}\enskip}
\newcommand{\ePf}{\medskip}
\def\QED{\hskip0.1em\hfill\null\ \null\nobreak\hfill\kern3pt\vbox{\hrule\hbox
   {\vrule\kern1pt\vbox{\kern1.7pt\hbox{$\scriptscriptstyle{QED}$}
    \kern0.2pt}\kern1pt\vrule}\hrule}}

\def\END{\hskip0.1em\hfill\null\ \null\nobreak\hfill\kern3pt\vbox{\hrule\hbox
   {\vrule\kern1pt\vbox{\kern1.7pt\hbox{$\,\,\,\vspace{5pt}$}
    \kern0.2pt}\kern1pt\vrule}\hrule}}
\newtheorem{theorem}{Theorem}
\newtheorem{lemma}{Lemma}
\newtheorem{corollary}{Corollary}
\newtheorem{proposition}{Proposition}
\newtheorem{remark}{Remark}
\newtheorem{definition}{Definition}
\newtheorem{example}{Example}
\newcommand{\owed}[1]{\overset{#1}{\wedge}}
\DeclareMathOperator{\byd}{{\raisebox{.1ex}{:}{=}}}
\newcommand{\bCd}{\bEq\begin{CD}}
\newcommand{\eCd}{\end{CD}\eEq}
\newcommand{\bcd}{\beq\begin{CD}}
\newcommand{\ecd}{\end{CD}\eeq}
\newcommand{\ben}{\begin{enumerate}}
\newcommand{\een}{\end{enumerate}}
\newcommand{\bEq}{\begin{eqnarray}}
\newcommand{\eEq}{\end{eqnarray}}
\newcommand{\beq}{\begin{eqnarray*}}
\newcommand{\eeq}{\end{eqnarray*}}
\newcommand{\bDf}{\begin{definition}\em}
\newcommand{\eDf}{\end{definition}}
\newcommand{\bLm}{\begin{lemma}}
\newcommand{\eLm}{\end{lemma}}
\newcommand{\bPr}{\begin{proposition}}
\newcommand{\ePr}{\end{proposition}}
\newcommand{\bTh}{\begin{theorem}}
\newcommand{\eTh}{\end{theorem}}
\newcommand{\bCr}{\begin{corollary}}
\newcommand{\eCr}{\end{corollary}}
\newcommand{\bRm}{\begin{remark}\em}
\newcommand{\eRm}{\end{remark}}
\newcommand{\bEx}{\begin{example}\em}
\newcommand{\eEx}{\end{example}}

\newcommand{\Z}{\mathbb{Z}}
\newcommand{\ie}{{\em i.e$.$} }
\newcommand{\eg}{{\em e.g$.$} }
\newcommand{\R}{I\!\!R}

\newcommand{\cE}{\mathcal{E}}

\newcommand{\cL}{\mathcal{L}}

\newcommand{\cS}{\mathcal{S}}

\newcommand{\bK}{\boldsymbol{K}}

\newcommand{\bR}{\boldsymbol{R}}

\newcommand{\bT}{\boldsymbol{T}}
\newcommand{\bU}{\boldsymbol{U}}

\newcommand{\bX}{\boldsymbol{X}}
\newcommand{\bY}{\boldsymbol{Y}}

\newcommand{\sub}{\subset}

\newcommand{\wed}{\wedge}
\newcommand{\com}{\!\circ\!}

\newcommand{\alp}{\alpha}
\newcommand{\bet}{\beta}

\newcommand{\del}{\delta}
\newcommand{\eps}{\epsilon}

\newcommand{\lam}{\lambda}
\newcommand{\sig}{\sigma}

\newcommand{\Gam}{\Gamma}

\DeclareFontFamily{U}{UWCyr}{}
\DeclareFontShape{U}{UWCyr}{m}{n}{%
    <5> <6> <7> <8> <9>
    <10> <10.95> <12> <14.4> <17.28> <20.74> <24.88> wncyr10
    }{}
\DeclareMathAlphabet{\cyrm}{U}{UWCyr}{m}{n}
\newcommand{\For}{{\Lambda}}
\newcommand{\Con}{{\mathcal{C}}}
\newcommand{\Hor}{{\mathcal{H}}}
\newcommand{\Var}{{\mathcal{V}}}
\newcommand{\Thd}{{\Theta}}
\renewcommand{\rfloor}{\con}
\def\con{{\offinterlineskip\lower 1truept\hbox{\kern2pt
\vbox to7truept{\vfill\hbox to4truept{\hrulefill}}\vrule \kern3pt}}}

\newcommand{\bdg}{\begin{diagram}}
\newcommand{\edg}{\end{diagram}}

\title{Variational Lie derivative and 
cohomology classes\footnote{Research  supported by the University of Torino {\em via} the research project {\em Teorie di campo classiche: calcolo delle variazioni etc.} and  partially by GNFM of INdAM; E.W. is also supported by MIUR research grant {\em  Sequenze variazionali e Teoremi di Noether etc.}.}}

\author{Marcella Palese and Ekkehart Winterroth \\ 
\footnotesize{Department of  Mathematics,
        University of Torino}\\
        \footnotesize{via C. Alberto 10, I-10123 Torino, Italy}\\
        \footnotesize{{\sc e-mail: [marcella.palese, ekkehart.winterroth]@unito.it}}}
\date{}

\begin{document}
\maketitle

\begin{abstract}
We relate cohomology defined by a system of local
Lagrangian with the cohomology class of the system of
local variational Lie derivative, which is in turn a local
variational problem; we show that the latter cohomology
class is zero, since  the variational Lie derivative
`trivializes' cohomology classes defined by variational
forms. As a consequence, conservation laws associated with
symmetries ensuring the vanishing of the second variational derivative of a local
variational problem  are globally defined.

\medskip


\end{abstract}

%-----------------------------------------------------------------------------%
\section{Introduction and preliminaries}
%-----------------------------------------------------------------------------%
 
The geometrical formulations of the calculus of variations
on fibered manifolds include a large class of theories
for which the Euler--Lagrange operator is a morphism of an exact
sequence \cite{AnDu80,Kru90,Tak79,Tul77,Vin77}. The module in degree $n+1$, consequently contains `equations', \ie dynamical form; the global inverse problems becomes then simple homological algebra: a given equation is an Euler--Lagrange equation if its dynamical form is the differential of a Lagrangian and this is equivalent to the `equation' being closed in the complex and its cohomolgy class being trivial, \ie Helmholtz conditions.
 `Equations' which are only {\em locally variational}, \ie which are closed in the complex and define a non trivial cohomology class admit a system of local Lagrangians, one for each open set in a suitable covering, 
which satisfy certain relations among them.
We shall consider generalized symmetries, \ie global projectable vector field on a jet fiber manifold which are symmetries of dynamical forms, in particular of locally variational dynamical forms.
This means that symmetries of the equations are chosen and corresponding formulations of Noether theorem (II) are considered in order to study obstruction to globality and relative properties of associated conserved quantities. 

We derive the local and global version of Noether theorems and the `inner' structure of the obstruction to the existence of global conserved currents that arises is investigated. 
We shall explicate relevant properties of  the {\em variational Lie derivative}, a differential operator acting on equivalence classes of variational forms in the variational sequence defined in \cite {FPV02}, and
relate the cohomology class defined by a system of local Lagrangian with the cohomology class defined by the system of local variational  Lie derivative, which is in turn a local variational problem. We show that the variational Lie derivative `trivializes' cohomology classes defined by variational forms.
The obstruction to the existence of a global conserved current for a local variational problem is the difference of two independent cohomology classes defined by means of  the variational Lie derivative. As a consequence we find that conservation laws associated with symmetries ensuring the vanishing of the {\em second variational derivative} of a local variational problem  are globally defined.

\medskip

We shall consider the variational sequence \cite{Kru90} defined on a fibered manifold $\pi : \bY \to \bX$, with $\dim
\bX = n$ and $\dim \bY = n+m$. For $r \geq 0$ we have
the $r$--jet space $J_r\bY$ of jet prolongations of sections of 
the fibered manifold $\pi$. We have also the natural fiberings $\pi^r_s : J_r\bY \to
J_s\bY$, $r \geq s$, and $\pi^r : J_r\bY \to \bX$; among these the
fiberings $\pi^r_{r-1}$ are {\em affine bundles} which  induce the natural fibered splitting \cite{Kru90} 
\beq 
J_r\bY\times_{J_{r-1}\bY}T^*J_{r-1}\bY \simeq
J_r\bY\times_{J_{r-1}\bY}(T^*\bX \oplus V^*J_{r-1}\bY)\,. 
\eeq
The above splitting induces also a decomposition of the exterior
differential on $\bY$ in the {\em horizontal\/} and {\em vertical
differential\/}, $(\pi^{r+1}_r)^*\com\, d = d_H + d_V$.
A {\em projectable vector field} on $\bY$ is defined to be a pair
$(\Xi,\xi)$, where the vector field $\Xi:\bY \to T\bY$ 
is a fibered morphism over
the vector field $\xi: \bX \to T\bX$. By $(j_{r}\Xi, \xi)$ we denote the jet
prolongation of $(\Xi,\xi)$, and by $j_{r}\Xi_{H}$ and
$j_{r}\Xi_{V}$ the horizontal and the vertical part
of $j_{r}\Xi$, respectively.

For $q \leq s $, we consider the standard sheaves $\For^{p}_{s}$
of $p$--forms on $J_s\bY$,
 the sheaves $\Hor^{p}_{(s,q)}$ and
$\Hor^{p}_{s}$ of {\em horizontal forms}, \ie of local {\em fibered morphisms} over
$\pi^{s}_{q}$ and $\pi^{s}$ of the type
$\alp : J_s\bY \to \owed{p}T^*J_q\bY$ and $\bet : J_s\bY \to \owed{p}T^*\bX$,
respectively. We also have the subsheaf $\Con^{p}_{(s,q)}
\sub \Hor^{p}_{(s,q)}$ of {\em contact forms\/}, \ie
of sections $\alp \in \Hor^{p}_{(s,q)}$ with values into
$\owed{p} (\Con^{*}_{q}[\bY])$.

According to \cite{Kru90}, the above fibered splitting yields the {\em sheaf splitting}
$\Hor^{p}_{(s+1,s)}$ $=$ $\bigoplus_{t=0}^p$
$\Con^{p-t}_{(s+1,s)}$ $\wed\Hor^{t}_{s+1}$, which restricts to the inclusion
$\For^{p}_s$ $\sub$ $\bigoplus_{t=0}^{p}$
$\Con^{p-t}{_s}\wed\Hor^{t,}{_{s+1}^{h}}$,
where $\Hor^{p,}{_{s+1}^{h}}$ $\byd$ $h(\For^{p}_s)$ for $0 < p\leq n$ and the map
$h$ is defined to be the restriction to $\For^{p}_{s}$ of the projection of
the above splitting onto the nontrivial summand with the highest
value of $t$.
Starting from this splitting one can define the sheaves of contact forms, \ie forms which do not have a variational role. 
In fact, let us denote by $d\ker h$ the sheaf generated by
the corresponding presheaf and set then $\Thd{*}_{r}$ $\equiv$ $\ker h$ $+$
$d\ker h$.
The quotient sequence
\beq
0\arrow{e} \R_{\bY} \arrow{e} \dots\,\
\arrow[4]{e,t}{\cE_{n-1}}\,\ \ \For^{n}_r/\Thd^{n}_r
\arrow[3]{e,t}{\cE_{n}}\,\ \For^{n+1}_r/\Thd^{n+1}_r
\arrow[4]{e,t}{\cE_{n+1}}\,\ \ \For^{n+2}_r/\Thd^{n+2}_r
\arrow[4]{e,t}{\cE_{n+2}}\,\ \ \dots\,\ \arrow{e,t}{d} 0
\eeq
defines the {\em $r$--th order variational sequence\/}
associated with the fibered manifold $\bY\to\bX$. It turns out
that it is an exact resolution of the constant sheaf $\R_{\bY}$
over $\bY$.

The quotient sheaves (the sections of which are classes of forms modulo contact forms) in the variational sequence can be represented as sheaves $\Var^{k}_{r}$ of $k$-forms on jet spaces of higher order. In particular, currents are classes  $\nu\in(\Var^{n-1}_{r})_{\bY}$;
Lagrangians are classes $\lam\in(\Var^{n}_{r})_{\bY}$, while $\cE_n(\lam)$ is called a Euler--Lagrange form (being 
$\cE_{n}$ the Euler--Lagrange morphism); dynamical forms are classes $\eta\in(\Var^{n+1}_{r})_{\bY}$ and $\cE_{n+1}(\eta) \byd \tilde{H}_{d\eta}$ is a Helmohltz form (being $\cE_{n+1}$ the corresponding Helmholtz morphism).

The cohomology groups of the corresponding  complex of global sections
\beq
0 \arrow{e} \R_{\bY} \arrow{e}  \dots\,\
\arrow[4]{e,t}{\cE_{n-1}}\,\ \ (\For^{n}_r/\Thd^{n}_r)_{\bY}
\arrow[3]{e,t}{\cE_{n}}\,\ \  (\For^{n+1}_r/\Thd^{n+1}_r)_{\bY}
\arrow[4]{e,t}{\cE_{n+1}}\,\ \  (\For^{n+2}_r/\Thd^{n+2}_r)_{\bY}
\arrow[4]{e,t}{\cE_{n+2}}\,\ \ \dots\,\
\arrow{e,t}{d} 0\,
\eeq will be denoted by $H^*_{\text{VS}}(\bY)$. 

For any sheaf  $\cS$ of
Abelian groups over a topological space $\mathfrak{T}$, and any  countable
open covering of $\mathfrak{T}$, denoted by  $\mathfrak{U}\equiv\{U_{i}\}_{i\in I}$,
with $I\sub \Z$, denote the set of $q$--cochains with
coefficients in $\cS$ by $C^{q}(\mathfrak{U},\cS )$.
 Let $\sig
=(U_{i_{0}},\dots ,U_{i_{q+1}})\sub \mathfrak{U}$ be a $q$--simplex and
$f\in C^{q}(\mathfrak{U},\cS )$.
The {\em coboundary operator} $\mathfrak{d}: C^{q}(\mathfrak{U},\cS )\to
C^{q+1}(\mathfrak{U},\cS )$ is the map defined by
$
\mathfrak{d} f(\sig )\equiv \sum^{q+1}_{i=0}(-1)^{i}r^{|\sig_{i}|}_{|\sig |}
f(\sig_{i})$, 
where $\sig_{j}\equiv (U_{i_{0}},\dots ,U_{i_{j-1}},U_{i_{j+1}},\dots
U_{i_{q+1}})$, for $0\leq j\leq q+1$, $r$ is the restriction mapping of
$\cS$ and $|\sig_{j}|$ denotes the lenght of $\sig_{j}$.
The coboundary $\mathfrak{d}$
is a group morphism, such that $\mathfrak{d} ^{2}=0$.
Hence we have the cochain complex
$
C^{0}(\mathfrak{U},\cS )\to C^{1}(\mathfrak{U},\cS )\to
C^{2}(\mathfrak{U},\cS )
\to\dots
$
The derived groups $H^{*}(\mathfrak{U},\cS )$ of the above cochain
complex are the {\em \v Cech cohomology} of the covering $\mathfrak{U}$ with
coefficients in $\cS$.  The {\em \v Cech cohomology} $H^{*}(\mathfrak{H},\cS )$ of $\mathfrak{T}$
with coefficients in $\cS$ is defined as the direct limit.

Since the variational sequence is a soft
resolution of the constant sheaf $\R_{\bY}$ over $\bY$,  the cohomology of the complex of global sections is naturally isomorphic to both the \v Cech cohomology of  $\bY$ with coefficients in the constant sheaf $\R$ 
and  the de Rham cohomology $ H^k_{\text{dR}}\bY$ \cite{Kru90}.

%-----------------------------------------------------------------------------%
\section{Local variational problems and cohomology}
%-----------------------------------------------------------------------------%

Let $\bK_{r}\byd \text{Ker}\,\, \cE_{n}$ and let 
$\cE_{n}(\Var^{n}_{r})$ be the sheave of Euler--Lagrange morphisms: for a global section $\eta\in(\Var ^{n+1}_{r})_{\bY}$ we have 
$\eta\in(\cE_{n}(\Var^{n}_{r}))_{\bY}$ if and only if $\cE_{n+1}(\eta)=0$, which are the Helmholtz conditons of local variationality.
A global inverse problem is to find necessary and sufficient conditions for such a locally variational $\eta$ to be globally variational.
We notice that the short exact sequence of sheaves
\beq 0
\arrow{e}\bK_{r}\arrow{e} \Var^{n}_{r}\arrow[2]{e,t}{\cE_{n}}\, \
\cE_{n}(\Var^{n}_{r})\arrow{e} 0 \eeq gives rise to the long exact
sequence in \v Cech cohomology \beq 0 \arrow{e} (\bK_{r})_{\bY} \arrow{e}
(\Var^{n}_{r})_{\bY} \arrow{e}
(\cE_{n}(\Var^{n}_{r}))_{\bY} \arrow{e,t}{\del} H^{1}(\bY,
\bK_{r})\arrow{e} 0 \,. 
\eeq 
Hence, every $\eta\in(\cE_{n}(\Var^{n}_{r}))_{\bY}$ (\ie locally variational) defines a cohomology class 
\beq \del [\eta] \in
H^{1}(\bY, \bK_{r}) \simeq H^{n+1}_{VS}(\bY)
\simeq H^{n+1}_{dR}(\bY)
\simeq H^{n+1}(\bY,\bR)\,.
\eeq
Analogously, let $\bT_{r}\byd \text{Ker}\,\, d_{H}$; the short exact sequence of sheaves
\beq 0 \to\bT_{r}\to
\Var^{n-1}_{r}\arrow{e,t}{d_{H}} d_{H}(\Var^{n-1}_{r})\to 0 \eeq
 gives rise to the long exact
sequence in \v Cech cohomology 
\beq 0 \to\Gam(\bY,\bT_{r})\to
\Gam(\bY,\Var^{n-1}_{r})\to
\Gam(\bY,d_{H}(\Var^{n-1}_{r}))\arrow{e,t}{\del '} H^{1}(\bY,
\bT_{r})\to 0 \,. \eeq 

Hence, every $\mu\in(d_H(\Var^{n-1}_{r}))_{\bY}$ (\ie variationally trivial) defines a cohomology class 
\beq \del' [\mu] \in
H^{1}(\bY,  \bT_{r}) \simeq H^{n}_{VS}(\bY)
\simeq H^{n}_{dR}(\bY)
\simeq H^{n}(\bY,\bR) \,.
\eeq
The solution to global inverse problem is now  simple and elegant: $\eta$ is globally variational if and only if $\del [\eta] = 0$, because only then there exists a global section  $\lam\in(\Var^{n}_{r})_{\bY}$ with $\eta = \cE_n(\lam)$. 
If instead $\del [\eta] \neq 0$ then $\eta =  \cE_n(\lam)$ can be solved only locally, \ie for any countable good covering of $\bY$, $\mathfrak{U}\equiv\{U_{i}\}_{i\in I}$,
 $I\sub \Z$,  there exist local Lagrangians $\lam_{i}$ over each subset $\bU_{i}\sub\bY$ such that $\eta_{i}=\cE_{n}(\lam_{i})$.
The local Lagrangians satisfy  $\cE_{n}((\lam_{i}-\lam_{j})|_{U_{i}\cap U_{j}}) = 0$ and conversely  any system of local sections with this property gives rise to an Euler--Lagrange form $\eta\in(\cE_{n}(\Var^{n}_{r}))_{\bY}$ with cohomology class  $\del [\eta] \in
H^{1}(\bY, \bK_{r})$. 

A system of local sections  $\lam_{i}$ of  $(\Var^{n}_{r})_{U_{i}}$ for an arbitrary covering  $\{\bU_{i}\}_{i\in \Z}$ in $\bY$ such that  $\cE_{n}((\lam_{i}-\lam_{j})|_{U_{i}\cap U_{j}}) = 0$, is what we call a {\em local variational problem}; two local variational problems are {\em equivalent } if and only if they give rise to the same Euler--Lagrange form. The covering  
$\mathfrak{U}$ of $\bY$ together with the local Lagrangians $\lam_{i}$ is called a {\em presentation} of the local variational problem \cite{FePaWi10}.
Note that every cohomology class in  $H^{n+1}_{dR}(\bY) \simeq H^{n+1}(\bY,\bR)$ gives  rise to local variational problems.
Non trivial  $H^{n+1}(\bY,\bR)$ can appear \eg when dealing with symmetry breaking, $\bY$ will then be fibred (over $\bX$) by homogeneous spaces. 
Two equivalent systems of local Lagrangians already defined with respect to the same covering can differ by an arbitrary $0$-cocycle of variationally trivial Lagrangians, \ie an arbitrary collections of local sections (over the $\bU_{i}\sub\bY$) of  $\bK_{r}$. In consequence, on a give open set,  the local Lagrangian from one system will have in general infinitesimal symmetries different from those of the local Lagrangian from the other. 

For any countable open covering of $\bY$,
$\lam= \{\lam_{i}\}_{i\in I}$ is then a $0$--cochain of Lagrangians in \v Cech
cohomology with values in the sheaf $\Var^{n}_{r}$, \ie $\lam\in
C^{0}(\mathfrak{U},\Var^{n}_{r})$. By an abuse of notation we shall
denote by $\eta_{\lam}$ the $0$--cochain formed by the restrictions
$\eta_{i}=\cE_{n}(\lam_{i})$.
Let $\mathfrak{d}\lam
=\{\lam_{ij}\}=(\lam_{i}-\lam_{j})|_{U_{i}\cap U_{j}}$. Of course, 
$\mathfrak{d}\lam =0$ if and only if $\lam$ is globally
defined on $\bY$; analogously, if $\eta \in
C^{0}(\mathfrak{U},\Var^{n+1}_{r})$, then $\mathfrak{d}\eta =0$ if
and only if $\eta$ is global.
Let $\lam\in C^{0}(\mathfrak{U},\Var^{n}_{r})$ and let
$\eta_{\lam}\equiv \cE_{n}(\lam)\in
C^{0}(\mathfrak{U},\Var^{n+1}_{r})$ be as above. We stress that
$\mathfrak{d}\lam=0$ implies $\mathfrak{d}\eta_{\lam}=0$, while by
$\R$--linearity  we have $\mathfrak{d}\eta_{\lam}=\eta_{\mathfrak{d}\lam}=0$ \ie
$\mathfrak{d}\lam\in C^{1}(\mathfrak{U},\bK_r)$ \cite{BFFP03}. 

%---------------------------------------------------------------------------
\subsection{Variational Lie derivative and cohomology classes}
%---------------------------------------------------------------------------

In order to formulate Noether theorems linking  symmetries of the local variational problem to conserved quantities, in \cite{FePaWi10} we tackled the question what the most natural choice for {\em symmetries of the local variational problem } might be.
We shall use extensively the concept of a {\em variational Lie derivative} operator $\cL_{j_{r}\Xi}$, defined for  any projectable vector field $(\Xi,\xi)$, which was inspired by the fact that the standard Lie derivative of forms with respect to a projectable vector field preserves the contact structure induced by the affine bundles $\pi^r_{r-1}$ (with $r\geq 1$) \cite{Kru73}. The variational Lie derivative is a local differential operator  by which symmetries  of Lagrangian and dynamical forms, and corresponding Noether theorems,  can be characterized; thus, we shall refer to Noether theorems as formulated in terms of variational Lie derivatives  of equivalence classes in the variational sequence \cite{FPV02}.

Let  $\eta_{\lam}$ be the Euler--Lagrange morphism of a local variational problem  and let $ \cL_{j_{r}\Xi}  \eta_{\lam}=0$.
Locally we have $\Xi_{V} \rfloor \eta_{\lam} = \Xi_{V} \rfloor \cE_{n}(\lam_{i})$.
The first Noether theorem implies that
$
0 $ $=$ $ \Xi_{V} \rfloor \eta_{\lam}  +
d_{H}( \eps_i (\lam_{i},\Xi) -  \beta(\lam_{i},\Xi) )$, where $\eps_i \byd\eps_i (\lam_{i},\Xi) = j_{r}\Xi_{V} \rfloor p_{d_{V}\lam_{i}}+ \xi \rfloor \lam_{i}$ is the usual {\em canonical} Noether current.
Along the solutions of Euler--Lagrange equations we thus get a {\em local conservation law}. We notice that in  \cite{BrKr05} local conserved currents are derived by using Lepagian equivalent of local systems of Lagrangians.
The conserved current  is $\eps(\lam_{i},\Xi) - \beta(\lam_{i},\Xi)$ which is a {\em local} object; in fact, 
 the Noether current $\eps(\lam_{i},\Xi)$ is conserved if and only if $\Xi$ is also a symmetry of $\lam_{i}$.
A local variational problem is a global object in the sense that it has a global Euler--Lagrange morphism defining a topological invariant. Consequently, there is also a precise relation between our local conservation laws \cite{FePaWi10}. 

In fact, let $\eta_{\lam}$ be the Euler--Lagrange morphism of a local variational problem and $\lam_{i}$ the system of local Lagrangians of an arbitrary given presentation.
The contraction $\Xi_{V} \rfloor \eta_{\lam}$ defines a cohomology class, since 
 $0=\cL_{j_{r}\Xi} \eta=
 \cE_{n}(\Xi_{V}\rfloor \eta_{\lam})$. Then  the local currents satisfy $d_{H}(\eps(\lam_{i},\Xi) - \beta(\lam_{i},\Xi) - \eps(\lam_{j},\Xi) + \beta(\lam_{j},\Xi)) = 0$. Thus we have  that
the local currents are the restrictions of  a global conserved current if and only if the cohomlogy class $ [\Xi_{V} \rfloor \eta_{\lam}] \in H^{n}_{dR}(\bY)$ vanishes.
It is noteworthy that also when the cohomolgy class $[ \cE_{n}(\lam)]$ is trivial, the cohomolgy class 
$[\Xi_{V} \rfloor \cE_{n}(\lam)]$ may be instead non trivial \cite{FePaWi10}.
\bRm
If $\Xi$ is a symmetry of all local Lagrangians $\lam_{i}$ of a given presentation of the local variational problem, the Noether currents are conserved and form a system of local potentials of the cohomology class $[\Xi_{V} \rfloor \eta_{\lam}] \in H^{n}_{dR}(\bY)$. 
In general, we have  
$d_{H}(\eps(\lam_{i},\Xi) - \eps(\lam_{j},\Xi))$ $=$ 
$\cL_{j_{r}\Xi}\lam _{i} -  \cL_{j_{r}\Xi}\lam _{j} \neq 0$,
thus neither the $\cL_{j_{r}\Xi}\lam _{i}$ nor the $d_{H}(\eps(\lam_{i},\Xi))$ are generally the restrictions of global closed $n$-forms. \END
\eRm

There are in  fact  two rather independent obstructions, one coming from the Lie derivative of the local Lagrangians being not necessarily zero, the other from the system of local Noether currents. 
In the cohomological obstruction to the existence of a global Lagrangian for a local variational problem, it is then of great interest to study how the  variational Lie derivative affects cohomology classes.

\bPr 
The variational Lie derivative transforms non trivial cohomology classes  to trivial cohomology classes associated with the variational Lie derivative of local presentations, \ie the variational Lie derivative `trivializes'  cohomology classes in the variational sequence.
\ePr
\bPf
In fact, by linearity we have
\beq
\eta_{\cL_{\Xi}\lam_{i}} = \cE_{n} (\Xi\rfloor \eta_{\lam}) + \cE_{n} (d_H \eps_i )=  \cE_{n} (\Xi\rfloor \eta_{\lam}) = \cL_{\Xi}\eta_{\lam}\,.
\eeq
Since $\cL_{\Xi}\eta_{\lam} =\cE_{n} (\Xi\rfloor \eta_{\lam})$ we have that $\del (\cL_{\Xi}\eta_{\lam})= \del(\eta_{\cL_{\Xi}\lam_{i}})  = 0$ although $\del(\eta_\lam) \neq 0$.  
\ePf

This result, by the way, holds true at any degree $k$ in the variational sequence and independently from the fact that $\Xi$ be a generalized symmetry or not.
In particular, from this we deduce the following.
\bCr
Euler--Lagrange equations of the local problem defined by $\cL_{\Xi}\lam_{i}$ are equal to Euler--Lagrange equations of the global problem defined by $\Xi\rfloor \eta_{\lam}$.
\eCr

In other words, the local problem defined by the local presentation $\cL_{\Xi}\lam_{i}$ is variationally equivalent to a global one.

The same can be stated for local variationally trivial Lagrangians. Suppose we have a global variationally trivial Lagrangian $\mu$, \ie such that $\cE_n(\mu )= 0$, this means that we have a 0-cocycle of currents $\nu_i$ such that $ \mu =d_H \nu_i$  and  $\mathfrak{d}\mu_\nu =0$ but we suppose $\del' (\mu_\nu)\neq 0$.
We can consider the Lie derivative $\cL_{\Xi}\nu_{i}$ and the corresponding $\mu_{\cL_{\Xi}\nu_{i}}$.
\bCr
Divergence equations (conditions for a current to be locally variationally trivial) of the local problem defined by $\cL_{\Xi}\nu_{i}$ are equal to divergence equations of the global problem defined by $\Xi_H\rfloor \mu_{\nu} +\Xi_V\rfloor p_{d_{V}\mu_{\nu}}$.
\eCr
\bPf
We have 
\beq
\mu_{\cL_{\Xi}\nu_{i}}=d_H (\Xi_H\rfloor \mu_{\nu} +\Xi_V\rfloor p_{d_{V}\mu_{\nu}})=  \cL_{\Xi}\mu_\nu \,,
\eeq
so that 
$\del '(\cL_{\Xi}\mu_{\nu})= \del'(\mu_{\cL_{\Xi}\nu_{i}})  = 0$,  although $\del'(\mu_\nu) \neq 0$.
\ePf

When $\Xi$ is a generalized symmetry we have $\cL_{\Xi}\eta_{\lam} 
 =0$ then, in particular, we have $\del (\cL_{\Xi}\eta_{\lam})\equiv 0$. 
Furthermore, if $\Xi$ is a generalized symmetry then $\cE_{n} (\Xi\rfloor \eta_{\lam}) =0$ then we have a $0$-cocycle $\nu_i$ as above,  defined by $ \mu_\nu = \Xi\rfloor \eta_{\lam}\byd d_H(\nu_i)$. 
We see that a way to find a system of global currents associated with a system of local currents is to take the Lie derivatives of the local system, for which we saw that we would have $\del '(\cL_{\Xi}\mu_{\nu})= \del '(\cL_{\Xi}(\Xi\rfloor \eta_{\lam})) = \del '(\cL_{\Xi}(d_H(\nu_i))  =  \del '(d_H(\cL_{\Xi}(\nu_i))=0$.

A natural question is now if there exist a way to find under which conditions such a variational Lie derivative of local currents is a system of {\em conserved} currents.
The answer to such a question involves Jacobi equations for the local system $\lam_i$.
\bPr
Let $\Xi$ be a symmetry of the Euler--Lagrange form $\eta_\lam$.
If the second variational derivative is vanishing, then we have the conservation law 
$d_H \cL_\Xi  (\nu_i+\eps_i)=0$, where $\cL_\Xi  (\nu_i+\eps_i)$ is a local representative of  a global conserved current.
\ePr
\bPf
In fact, let us apply twice the variational Lie derivative; since we are supposing that $\Xi$ is a {\em generalized symmetry}, we have
\beq
\cL_\Xi \cL_\Xi \lam_i = \cL_\Xi  (\Xi \rfloor \eta_\lam) + \cL_\Xi (d_H (\eps_i)) = \cL_\Xi  (d_H(\nu_i)) +\cL_\Xi (d_H (\eps_i))= d_H \cL_\Xi  (\nu_i+\eps_i)\,.
\eeq
Thus the statement is an immediate consequence of the condition
$\cL_\Xi \cL_\Xi \lam_i \byd \del^2  \lam_i =0\,.
$
\ePf
Notice that 
$\cL_\Xi \cL_\Xi \lam_i =0$ means that the generalized symmetry $\Xi$ is required to be a generalized symmetry {\em and} also a symmetry of the local variational problem $\cL_\Xi \lam_i$.

%---------------------------------------------

\end{document}